\documentclass[amsmath,aps,prd,showpacs,a4paper,10pt]{revtex4}

 \usepackage{epsf}
 \usepackage{graphicx}    

\begin{document}

 \newcommand{\be}[1]{\begin{equation}\label{#1}}
 \newcommand{\ee}{\end{equation}}
 \newcommand{\bea}{\begin{eqnarray}}
 \newcommand{\eea}{\end{eqnarray}}
 \def\disp{\displaystyle}

 \def\gsim{ \lower .75ex \hbox{$\sim$} \llap{\raise .27ex \hbox{$>$}} }
 \def\lsim{ \lower .75ex \hbox{$\sim$} \llap{\raise .27ex \hbox{$<$}} }

 \begin{titlepage}

 \begin{flushright}
 arXiv:0803.3292
 \end{flushright}

 \title{\Large \bf How to Distinguish Dark Energy and
 Modified Gravity?}

 \author{Hao~Wei\,}
 \email[\,email address:\ ]{haowei@mail.tsinghua.edu.cn}
 \affiliation{Department of Physics and Tsinghua Center for
 Astrophysics,\\ Tsinghua University, Beijing 100084, China}

 \author{Shuang~Nan~Zhang}
 \affiliation{Department of Physics and Tsinghua Center for
 Astrophysics,\\ Tsinghua University, Beijing 100084, China\\
 Key Laboratory of Particle Astrophysics, Institute of High
 Energy Physics,\\
 Chinese Academy of Sciences, Beijing 100049, China\\
 Physics Department, University of Alabama in Huntsville,
 Huntsville, AL 35899, USA}

 \begin{abstract}\vspace{1cm}
 \centerline{\bf ABSTRACT}\vspace{2mm}
The current accelerated expansion of our universe could be due
 to an unknown energy component (dark energy) or a modification
 to general relativity (modified gravity). In the literature,
 it has been proposed that combining the probes of the cosmic
 expansion history and growth history can distinguish between
 dark energy and modified gravity. In this work, without
 invoking non-trivial dark energy clustering, we show that the
 possible interaction between dark energy and dark matter
 could make the interacting dark energy model and the modified
 gravity model indistinguishable. An explicit example is also
 given. Therefore, it is required to seek some complementary
 probes beyond the ones of cosmic expansion history and growth
 history.
 \end{abstract}


 \maketitle

 \end{titlepage}

 \renewcommand{\baselinestretch}{1.6}



\section{Introduction}\label{sec1}
The current accelerated expansion of our
 universe~\cite{r1,r2,r3,r4,r5,r6,r7,r8,r9,r10} has been one of the
 most active fields in modern cosmology. There are very strong
 model-independent evidences~\cite{r11} (see also e.g.~\cite{r12})
 for the accelerated expansion. Many cosmological models have been
 proposed to interpret this mysterious phenomenon, see
 e.g.~\cite{r1} for a recent review.

In the flood of various cosmological models, one of the most
 important tasks is to distinguish between them. The current
 accelerated expansion of our universe could be due to an unknown
 energy component (dark energy) or a modification to general
 relativity (modified gravity)~\cite{r1,r13}. Recently, some
 efforts have focused on differentiating dark energy and modified
 gravity with the growth function
 $\delta(z)\equiv\delta\rho_m/\rho_m$ of the linear matter density
 contrast as a function of redshift $z$. Up until now, most
 cosmological observations merely probe the expansion history
 of our universe~\cite{r1,r2,r3,r4,r5,r6,r7,r8,r9,r10}. As is
 well known, it is very easy to build models which share the same
 cosmic expansion history by means of reconstruction between models.
 Therefore, to distinguish various models, some independent and
 complementary probes are required. Recently, it was argued that the
 measurement of growth function $\delta(z)$ might be competent, see
 e.g.~\cite{r13,r14,r15,r16,r17,r18,r19,r20,r21,r22,r23,r24,
 r25,r47,r48,r49,r50}. If the dark energy model and modified
 gravity model share the same cosmic expansion history, they
 might have {\em different} growth histories. Thus, they might be
 distinguished from each other.

However, the approach mentioned above has been challenged by some
 authors. For instance, in~\cite{r26}, Kunz and Sapone explicitly
 demonstrated that the dark energy models with {\em non-vanishing
 anisotropic stress} cannot be distinguished from modified gravity
 models (e.g. DGP model) using growth function. In~\cite{r27},
 Bertschinger and Zukin found that if dark energy is generalized to
 include {\em both entropy and shear stress perturbations}, and the
 dynamics of dark energy is unknown {\it a priori}, then modified
 gravity [e.g. $f(R)$ theories] cannot in general be distinguished
 from dark energy using cosmological linear perturbations.

Here, we investigate this issue in another way. Instead of
 invoking non-trivial dark energy clustering (e.g.
 non-vanishing anisotropic stress), it is of interest to see
 whether the interaction between dark energy and cold dark
 matter can make dark energy and modified gravity
 indistinguishable. In fact, since the nature of both dark
 energy and dark matter are still unknown, there is no physical
 argument to exclude the possible interaction between them. On
 the contrary, some observational evidences of this interaction
 have been found recently. For example, in a series of papers
 by Bertolami {\it et al.}~\cite{r28}, they show that the Abell
 Cluster A586 exhibits evidence of the interaction between dark
 energy and dark matter, and they argue that this interaction might
 imply a violation of the equivalence principle. On the other hand,
 in~\cite{r29}, Abdalla {\it et al.} found the signature of
 interaction between dark energy and dark matter by using optical,
 X-ray and weak lensing data from 33 relaxed galaxy clusters.
 In~\cite{r42}, Ichiki and Keum discussed the cosmological
 signatures of interaction between dark energy and massive neutrino
 (which is also a candidate for dark matter) by using CMB power
 spectra and matter power spectrum. Therefore, it is reasonable to
 consider the interaction between dark energy and dark matter in
 cosmology. Since dark energy can decay into cold dark matter
 (and vice versa) through interaction, both expansion history and
 growth history can be affected simultaneously, similar to the case
 of modified gravity. Thus, it is natural to ask whether the
 combined probes of expansion history and growth history can
 distinguish between interacting dark energy models and modified
 gravity models. The answer might be no. So, it is required to seek
 some complementary probes beyond the ones of cosmic expansion
 history and growth history.

In this note, we propose a general approach in Sec.~\ref{sec2} to
 build an interacting quintessence model which shares both the same
 expansion history and growth history with the modified gravity
 model. In Sec.~\ref{sec3}, following this prescription, as an
 example, we explicitly demonstrate the interacting quintessence
 model which is indistinguishable with the DGP model in this sense.
 A brief discussion is given in Sec.~\ref{sec4}, in which some
 suggestions to break this degeneracy are also discussed.


\section{General formalism}\label{sec2}
Quintessence~\cite{r30,r31} is a well-known dark energy candidate.
 In this section, we consider the interacting quintessence
 model~\cite{r32,r33,r34,r35,r36,r37} and show how it can share
 both the same expansion history and growth history with modified
 gravity models, without invoking non-trivial dark energy clustering.

We consider a flat Friedmann-Robertson-Walker (FRW) universe. As is
 well known, the quintessence is described by a canonical scalar
 field with a Lagrangian density
 ${\cal L}_\phi=\frac{1}{2}\left(\partial_{\mu}\phi\right)^2-V(\phi)$.
 Assuming the scalar field $\phi$ is homogeneous, one obtains the
 pressure and energy density for quintessence
 \be{eq1}
 p_\phi=\frac{1}{2}\dot{\phi}^2-V(\phi),~~~~~~~
 \rho_\phi=\frac{1}{2}\dot{\phi}^2+V(\phi),
 \ee
 where a dot denotes the derivative with respect to cosmic time $t$.
 The Friedmann equation reads
 \be{eq2}
 3H^2=\kappa^2\left(\rho_m+\rho_\phi\right),
 \ee
 where $H\equiv\dot{a}/a$ is the Hubble parameter; $a=(1+z)^{-1}$
 is the scale factor (we have set $a_0=1$; the subscript ``0''
 indicates the present value of corresponding quantity; $z$ is the
 redshift); $\rho_m$ is the energy density of cold dark matter (we
 assume the baryon component to be negligible);
 $\kappa^2\equiv 8\pi G$. We assume that quintessence and cold dark
 matter interact through~\cite{r32,r33,r34}
 \bea
 &&\dot{\rho}_m+3H\rho_m=-\kappa Q\rho_m\dot{\phi},\label{eq3}\\
 &&\dot{\rho}_\phi+3H\left(\rho_\phi+p_\phi\right)
 =\kappa Q\rho_m\dot{\phi},\label{eq4}
 \eea
 which preserves the total energy conservation equation
 $\dot{\rho}_{tot}+3H\left(\rho_{tot}+p_{tot}\right)=0$. The
 dimensionless coupling coefficient $Q=Q(\phi)$ is an arbitrary
 function of $\phi$. Eq.~(\ref{eq4}) is equivalent to
 \be{eq5}
 \ddot{\phi}+3H\dot{\phi}+\frac{dV}{d\phi}=\kappa Q\rho_m.
 \ee
 Using Eqs.~(\ref{eq2})---(\ref{eq4}), one can obtain the
 Raychaudhuri equation
 \be{eq6}
 \dot{H}=-\frac{\kappa^2}{2}\left(\rho_m+\rho_\phi+p_\phi\right)
 =-\frac{\kappa^2}{2}\left(\rho_m+\dot{\phi}^2\right).
 \ee
 It is worth noting that due to the non-vanishing interaction,
 $\rho_m$ does not scale as $a^{-3}$. The above equations are
 associated with the expansion history. On the side of growth
 history, as shown in~\cite{r33}, the perturbation equation in
 the sub-horizon regime is
 \be{eq7}
 \delta^{\prime\prime}+\left(2+\frac{H^\prime}{H}
 -\kappa Q\phi^\prime\right)\delta^\prime
 =\frac{3}{2}\left(1+2Q^2\right)\Omega_m\delta,
 \ee
 where $\delta\equiv\delta\rho_m/\rho_m$ is the linear matter
 density contrast; $\Omega_m\equiv\kappa^2\rho_m/(3H^2)$ is the
 fractional energy density of cold dark matter; and a prime denotes
 a derivative with respect to $\ln a$. Note that in~\cite{r33} the
 absence of anisotropic stress has been assumed, namely, in
 longitudinal (conformal Newtonian) gauge the metric perturbations
 $\Phi=\Psi$. Obviously, when $Q=0$, Eq.~(\ref{eq7}) reduces to
 the standard form in general
 relativity~\cite{r14,r15,r16,r17,r18,r24,r38}
 \be{eq8}
 \ddot{\delta}+2H\dot{\delta}=4\pi G\rho_m\delta.
 \ee
 In fact, Eq.~(\ref{eq7}) from~\cite{r33} is valid
 for any $Q=Q(\phi)$ and generalizes the one of~\cite{r34} which
 is only valid for constant $Q$. On the other hand, in modified
 gravity, the perturbation equation~(\ref{eq8}) has been modified
 to~\cite{r18,r21,r22,r38,r39}
 \be{eq9}
 \ddot{\tilde{\delta}}+2\tilde{H}\dot{\tilde{\delta}}
 =4\pi G_{\rm eff}\rho_m\tilde{\delta},
 \ee
 where the quantities in modified gravity are labeled by a tilde
 ``$\sim$''; $G_{\rm eff}$ is the effective local gravitational
 ``constant'' measured by Cavendish-type experiment, which is
 time-dependent. In general, $G_{\rm eff}$ can be written as
 \be{eq10}
 G_{\rm eff}=G\left(1+\frac{1}{3\beta}\right),
 \ee
 where $\beta$ is determined once we specify the modified gravity
 theory. Eq.~(\ref{eq9}) can be rewritten as
 \be{eq11}
 \tilde{\delta}^{\prime\prime}+\left(2
 +\frac{\tilde{H}^\prime}{\tilde{H}}\right)\tilde{\delta}^\prime
 =\frac{3}{2}\left(1+\frac{1}{3\beta}\right)
 \tilde{\Omega}_m\tilde{\delta}.
 \ee

Now, we require that the interacting quintessence model shares
 both the same expansion history and growth history with the
 modified gravity model. That is, we identify
 \be{eq12}
 H=\tilde{H}~~~{\rm and}~~~\delta=\tilde{\delta}.
 \ee
 Comparing Eq.~(\ref{eq7}) with Eq.~(\ref{eq11}), we find that
 \be{eq13}
 \kappa Q\phi^\prime\delta^\prime=\frac{3}{2}\delta\left[
 \left(1+\frac{1}{3\beta}\right)\tilde{\Omega}_m
 -\left(1+2Q^2\right)\Omega_m\right].
 \ee
 Note that $\Omega_m\not=\tilde{\Omega}_m$ in general. From
 Eq.~(\ref{eq6}), we have
 \be{eq14}
 \left(\kappa\phi^\prime\right)^2=-3\Omega_m-2\frac{H^\prime}{H}.
 \ee
 We can recast Eq.~(\ref{eq3}) to
 \be{eq15}
 \Omega_m^\prime=-\left(3+2\frac{H^\prime}{H}
 +\kappa Q\phi^\prime\right)\Omega_m.
 \ee
 It turns out
 \be{eq16}
 \kappa Q\phi^\prime=-3-2\frac{H^\prime}{H}
 -\frac{\Omega_m^\prime}{\Omega_m}.
 \ee
 From Eqs.~(\ref{eq14}) and~(\ref{eq16}), we obtain
 \be{eq17}
 Q^2=\frac{\left(\kappa Q\phi^\prime\right)^2}
 {\left(\kappa\phi^\prime\right)^2}=\frac{\left(3+\disp
 2\frac{H^\prime}{H}+\frac{\Omega_m^\prime}{\Omega_m}\right)^2}
 {\disp -3\Omega_m-2\frac{H^\prime}{H}}.
 \ee
 Noting that $\delta=\tilde{\delta}$, we can find $\delta$ in
 Eq.~(\ref{eq13}) from Eq.~(\ref{eq11}). Once $\tilde{\Omega}_m$,
 $\beta$, $\tilde{H}$, and corresponding $\tilde{\delta}$ in the
 modified gravity are given, substituting Eqs.~(\ref{eq16})
 and~(\ref{eq17}) into Eq.~(\ref{eq13}) and noting Eq.~(\ref{eq12}),
 we obtain a differential equation for $\Omega_m$ with respect
 to $\ln a$. After we find $\Omega_m(\ln a)$ from this differential
 equation, $\kappa\phi^\prime(\ln a)$ can be obtained from
 Eq.~(\ref{eq14}), while $H=\tilde{H}$. Then, by using
 Eqs.~(\ref{eq14}) and~(\ref{eq16}), $Q(\ln a)=\left(\kappa
 Q\phi^\prime\right)/\left(\kappa\phi^\prime\right)$ is in hand.
 From Eq.~(\ref{eq2}), $\Omega_\phi\equiv\kappa^2\rho_\phi/(3H^2)
 =1-\Omega_m$. Noting Eq.~(\ref{eq1}) and using Eq.~(\ref{eq14}),
 we find the dimensionless potential of quintessence
 \be{eq18}
 U\equiv\frac{\kappa^2 V}{H_0^2}=3E^2\left(1-\frac{\Omega_m}{2}
 +\frac{1}{3}\frac{H^\prime}{H}\right),
 \ee
 where $E\equiv H/H_0$. Notice that $H^\prime/H=E^\prime/E$. By
 integrating $\kappa\phi^\prime(\ln a)$, we find $\kappa\phi$
 as a function of $\ln a$. Therefore, we can finally obtain
 $Q$, $U$ as functions of $\kappa\phi$.

In fact, this general approach proposed here is just a normal
 reconstruction method. For {\em any} given modified gravity
 model, we can {\em always} construct an interacting
 quintessence model in the framework of general relativity
 which shares both the same expansion history and growth
 history with this given modified gravity model. As is well known,
 an interacting quintessence model can be completely described
 by its potential $V(\phi)$ and the coupling $Q(\phi)$. For a
 given modified gravity model, its $\tilde{\Omega}_m$, $\beta$,
 and $\tilde{H}$ are known, whereas its corresponding
 $\tilde{\delta}$ can be obtained from Eq.~(\ref{eq11}).
 Following the procedure described in the text below
 Eq.~(\ref{eq17}), one can easily reconstruct the dimensionless
 potential $U(\kappa\phi)$ [which is equivalent to the
 potential $V(\phi)$ obviously] and the coupling
 $Q(\kappa\phi)$ [which is $Q(\phi)$ in fact] for the
 interacting quintessence model. Up until now, the desired
 interacting quintessence model which shares both the same
 expansion history and growth history with the given modified
 gravity model has been constructed. Therefore, the
 cosmological observations might be unable to distinguish
 between them, unless other complementary probes beyond the
 ones of cosmic expansion history and growth history are used.


 \begin{center}
 \begin{figure}[htbp]
 \centering
 \includegraphics[width=0.85\textwidth]{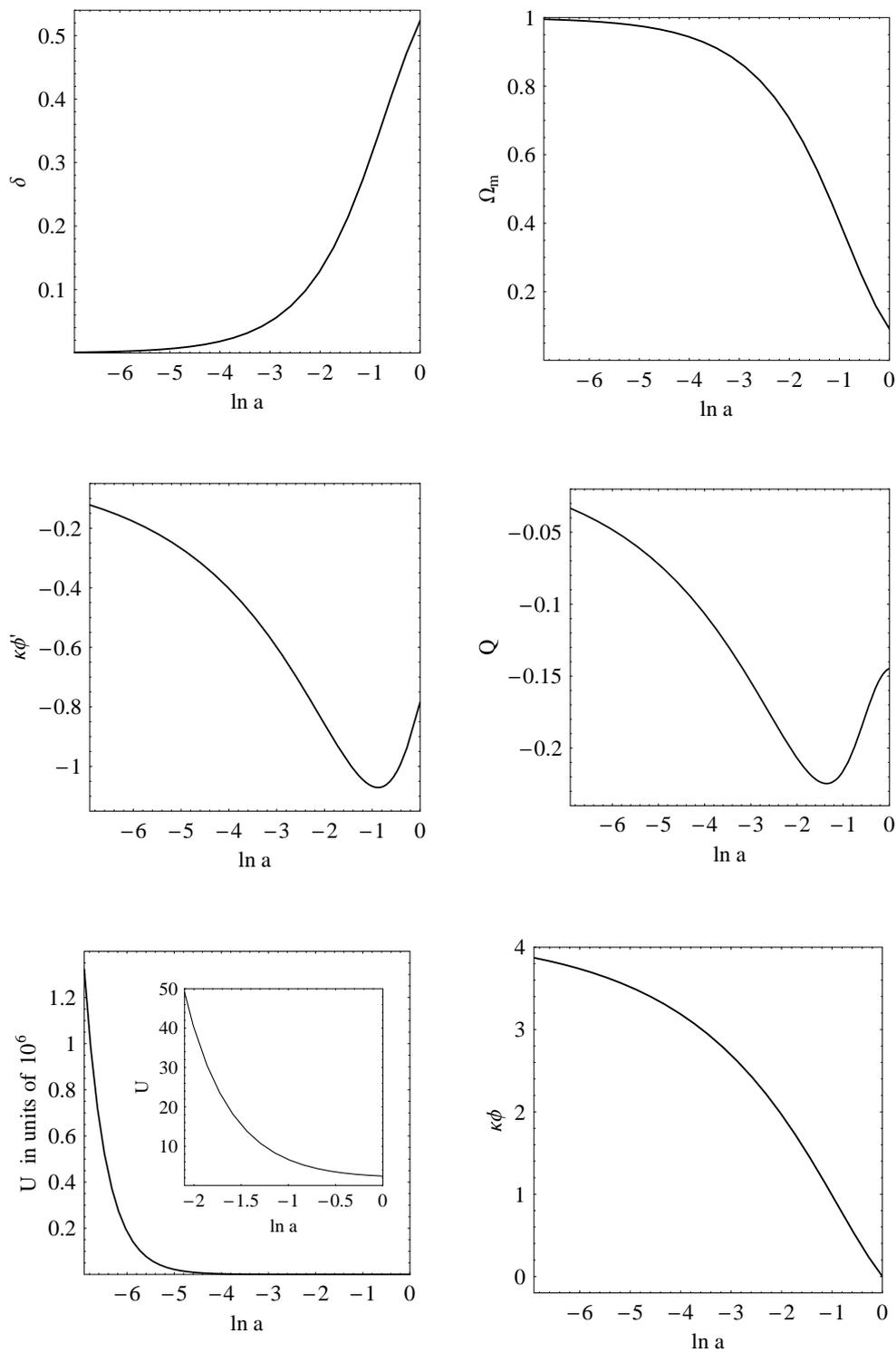}
 \caption{\label{fig1}
 $\delta=\tilde{\delta}$, $\Omega_m$, $\kappa\phi^\prime$,
 $Q$, $U\equiv\kappa^2 V/H_0^2$, and $\kappa\phi$
 as functions of $\ln a$. See text for details.}
 \end{figure}
 \end{center}



\section{Explicit example}\label{sec3}
In this section, we give an explicit example following the
 prescription proposed in Sec.~\ref{sec2}. Here, we consider
 the DGP braneworld model~\cite{r40} (see also
 e.g.~\cite{r21,r22,r18,r41}), which is the simplest modified
 gravity model. Assuming the flatness of our universe, in the
 DGP model (here we only consider the self-accelerating
 branch), $\tilde{E}\equiv\tilde{H}/\tilde{H}_0$
 is given by~\cite{r18,r21,r22}
 \be{eq19}
 \tilde{E}=\sqrt{\tilde{\Omega}_{m0}(1+z)^3+\tilde{\Omega}_{r_c}}+
 \sqrt{\tilde{\Omega}_{r_c}}=\sqrt{\tilde{\Omega}_{m0}\,e^{-3\ln a}
 +\tilde{\Omega}_{r_c}}+\sqrt{\tilde{\Omega}_{r_c}}\,,
 \ee
 where $\tilde{\Omega}_{r_c}$ is constant. $\tilde{E}(z=0)=1$
 requires
 \be{eq20}
 \tilde{\Omega}_{m0}=1-2\sqrt{\tilde{\Omega}_{r_c}}\,.
 \ee
 Therefore, the DGP model has only one independent model parameter.
 The fractional energy density of matter in the DGP model reads
 \be{eq21}
 \tilde{\Omega}_m=\frac{\tilde{\Omega}_{m0}(1+z)^3}{\tilde{E}^2(z)}
 =\frac{\tilde{\Omega}_{m0}\,e^{-3\ln a}}{\tilde{E}^2(\ln a)}.
 \ee
 In addition, the $\beta$ in Eq.~(\ref{eq10}) for the flat DGP
 model is given by~\cite{r18,r21,r22}
 \be{eq22}
 \beta=-\frac{1+\tilde{\Omega}_m^2}{1-\tilde{\Omega}_m^2}.
 \ee

\begin{center}
 \begin{figure}[htbp]
 \centering
 \includegraphics[width=0.9\textwidth]{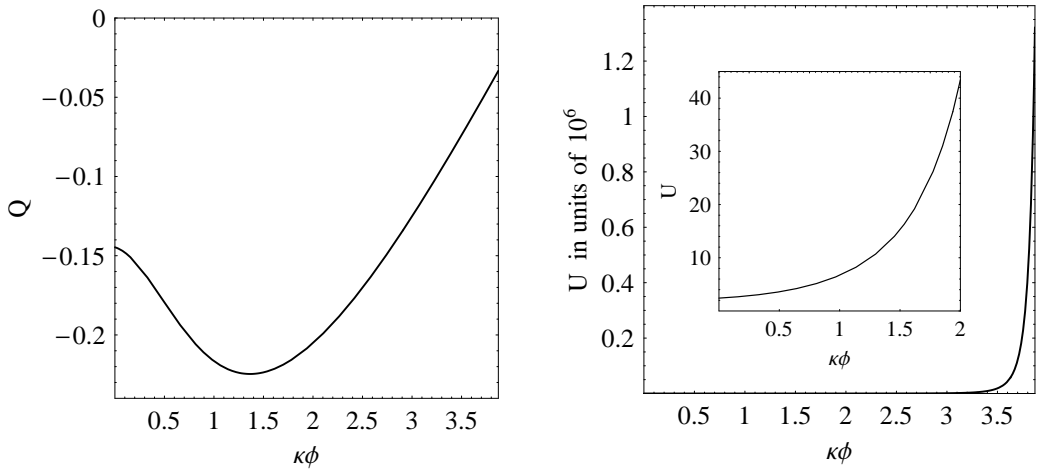}
 \caption{\label{fig2}
 $Q$ and $U\equiv\kappa^2 V/H_0^2$ as functions of $\kappa\phi$.}
 \end{figure}
 \end{center}

Fitting the DGP model to the 192 SNIa data compiled by
 Davis {\it et al.}~\cite{r10} which are joint data from
 ESSENCE~\cite{r9} and Gold07~\cite{r3}, we find that the best-fit
 parameter $\tilde{\Omega}_{r_c}=0.170$ with $\chi^2_{min}=196.128$
 while the corresponding $\tilde{\Omega}_{m0}$ is given by
 Eq.~(\ref{eq20}). Substituting this $\tilde{\Omega}_{m0}$ into
 Eqs.~(\ref{eq21}), (\ref{eq22}) and~(\ref{eq19}),
 $\tilde{\Omega}_m$, $\beta$, and $\tilde{E}$ as functions of
 $\ln a$ are known (since our main aim is to demonstrate the
 prescription proposed in Sec.~\ref{sec2}, we do not consider
 the errors, even one can use any $\tilde{\Omega}_{m0}$ here
 for demonstration). Noting that $H^\prime/H=E^\prime/E$,
 following the prescription proposed in Sec.~\ref{sec2}, we can
 easily construct the desired interacting quintessence model which
 shares both the same expansion history and growth history with the
 corresponding DGP model. At the first step, we obtain
 $\delta=\tilde{\delta}$ from Eq.~(\ref{eq11}). As is well known,
 $\tilde{\delta}^\prime=\tilde{\delta}=a$ at $z\gg 1$
 (see e.g.~\cite{r14,r15}). Thus, we use the initial
 condition $\tilde{\delta}^\prime=\tilde{\delta}=a_{ini}$ at
 $z_{ini}=1000$ for the differential equation~(\ref{eq11}). The
 resulting $\delta=\tilde{\delta}$ as a function of $\ln a$ is
 shown in Fig.~\ref{fig1}. At the second step, substituting
 Eqs.~(\ref{eq16}) and~(\ref{eq17}) into Eq.~(\ref{eq13}), we
 obtain $\Omega_m$ as a function of $\ln a$ from the resulting
 differential equation. Note that cold dark matter can decay to
 quintessence through interaction, $\Omega_m$ is unnecessary to
 be 1 at high redshift. So, for demonstration, we choose the
 initial condition $\Omega_m(z=z_{ini})=0.995$ at $z_{ini}=1000$
 for the differential equation of $\Omega_m$. Different values of
 $\Omega_m(z=z_{ini})$ only mean different displacements of
 $\kappa\phi^\prime$ and the dimensionless potential
 $U\equiv\kappa^2 V/H_0^2$ at $z_{ini}$. The resulting $\Omega_m$
 as a function of $\ln a$ is also shown in Fig.~\ref{fig1}. Then,
 following the prescription proposed in Sec.~\ref{sec2}, it is
 straightforward to obtain $\kappa\phi^\prime$, $Q$,
 $U\equiv\kappa^2 V/H_0^2$, and $\kappa\phi$ as functions of
 $\ln a$, while for demonstration we choose the negative branch
 for $\kappa\phi^\prime$, and choose $\phi_0=0$ when we get
 $\kappa\phi$. They are also shown in Fig.~\ref{fig1}. Once we
 obtain $Q$, $U$ and $\kappa\phi$ as functions of $\ln a$, it is
 easy to find $Q$ and $U$ as functions of $\kappa\phi$. The results
 are shown in Fig.~\ref{fig2}.

In short, here we have faithfully followed the reconstruction method
 proposed in Sec.~\ref{sec2}, and successfully constructed an
 interacting quintessence model which shares both the same expansion
 history and growth history with the DGP model whose single
 model parameter $\tilde{\Omega}_{r_c}=0.170$ which is the best-fit
 to the 192 SNIa data compiled by Davis {\it et al.}~\cite{r10}
 for example. The potential $V(\phi)$ and coupling $Q(\phi)$ which
 are required to describe the reconstructed interacting quintessence
 model have been presented in Fig.~\ref{fig2} through the equivalent
 $U(\kappa\phi)$ and $Q(\kappa\phi)$. There are no analytical
 expressions for them and instead the reconstructed $U(\kappa\phi)$,
 $Q(\kappa\phi)$ are given in numerical forms.


\section{Conclusion and discussion}\label{sec4}
In summary, we proposed a general approach to build an interacting
 quintessence model which shares both the same expansion history
 and growth history with the modified gravity model. Therefore, the
 cosmological observations might be unable to distinguish between
 them, unless other complementary probes beyond the ones of cosmic
 expansion history and growth history are used. As an example,
 we also explicitly demonstrated the interacting quintessence model
 which is indistinguishable with the DGP model in this sense.
 In fact, this proposed prescription is also valid for other
 modified gravity models, such as $f(R)$ theories, braneworld-type
 models, scalar-tensor theories (including Brans-Dicke theory),
 TeVeS/MOND models, and so on~\cite{r43,r44,r45}. Of course,
 one can also extend the interacting quintessence model to
 other interacting dark energy models.


In this note, without invoking non-trivial dark energy
 clustering (e.g. non-vanishing anisotropic stress), we find
 that the interaction between dark energy and dark matter could
 also make dark energy model and modified gravity model
 indistinguishable. How can we break this degeneracy? Firstly,
 we should carefully check the evidences of the interaction
 between dark energy and dark matter~\cite{r28,r29,r42}. If
 this interaction does not exist, the combined probes of cosmic
 expansion history and growth history might be enough to
 distinguish between dark energy and modified gravity. It is
 worth noting that the current observational constraints on the
 interaction between dark energy and dark matter in the
 literature (e.g.~\cite{r28,r29,r42,r46}) considered other
 interaction forms (e.g. $\propto H\rho_m$, $H\rho_{de}$ or
 $H\rho_{tot}$) which are different from the one of
 the present work [cf. Eqs.~(\ref{eq3}) and~(\ref{eq4})], and
 hence cannot be used to compare with the interaction
 considered here. Therefore, it is of interest to consider the
 observational constraints on the type of interactions
 $\propto Q(\phi)\rho_m\dot{\phi}$ in future works. Secondly,
 to break the degeneracy, the other complementary probes
 beyond the ones of cosmic expansion history and growth history
 are desirable. For instance, these complementary probes might
 include local tests of gravity, high energy phenomenology, and
 non-linear structure formation. Thirdly, in addition to the
 linear matter density contrast $\delta(z)$, one can also test
 the metric perturbations $\Phi$ and $\Psi$ by using the
 relationship between gravitational lensing and matter
 overdensity~\cite{r50}. For the interacting quintessence model
 $\Phi=\Psi$, whereas for the DGP model $\Phi\not=\Psi$. Thus,
 it is possible to distinguish between them by using delicate
 measurements. We consider that this issue deserves further
 investigations and believe that a promising future is
 awaiting us.


\section*{ACKNOWLEDGMENTS}
We thank the anonymous referee for comments, which helped us to
 improve this manuscript. We are grateful to
 Prof.~Rong-Gen~Cai, Prof.~Pengjie~Zhang, Prof.~Bin~Wang and
 Prof.~Xinmin~Zhang for helpful discussions and comments. We
 also thank Minzi~Feng, as well as Nan~Liang, Rong-Jia~Yang,
 Wei-Ke~Xiao, Pu-Xun~Wu, Jian~Wang, Lin~Lin, Bin~Fan and Yuan~Liu,
 for kind help and discussions. The major part of this work was
 completed during March 16--23, 2008. We acknowledge partial
 funding support from China Postdoctoral Science Foundation,
 and by the Ministry of Education of China, Directional
 Research Project of the Chinese Academy of Sciences under
 Project No.~KJCX2-YW-T03, and by the National Natural Science
 Foundation of China under Project No.~10521001.

 \renewcommand{\baselinestretch}{1.1}


\end{document}